\documentclass[aps,prl,reprint,nofootinbib,twoside,superscriptaddress,longbibliography,floatfix]{revtex4-2}
\usepackage[english]{babel}
\usepackage[utf8]{inputenc}
\usepackage{amsfonts,amssymb,amsmath}
\usepackage{mathtools}
\usepackage{amsthm}
\usepackage{bm}

\usepackage{tikz}
\usetikzlibrary{arrows.meta,positioning,calc}
\usepackage{xcolor}
\usepackage{hyperref}

\newcommand{\figlink}[2]{\hyperref[#1]{\textcolor{blue}{#2}}}

\usepackage[normalem]{ulem}
\hypersetup{
  colorlinks=true,
  citecolor=blue!55!black,
  linkcolor=blue!55!black,
  urlcolor=blue!55!black,
  pdftitle={More mutually unbiased bases},
  pdfauthor={Mateo Cárdenes Wuttig and Joseph Tindall}
}

\theoremstyle{definition}

\newif\ifdraft
\drafttrue 

\usepackage{booktabs}
\newcommand{\C}{\mathbb C}

\newcommand{\ket}[1]{\left|#1\right\rangle}

\newcommand{\diag}{\operatorname{diag}}

\begin{document}

\title{More mutually unbiased bases} 
\author{Mateo C\'ardenes Wuttig}
\thanks{Contact author: \href{mailto:mateo.cardeneswuttig@yale.edu}%
{mateo.cardeneswuttig@yale.edu}}
\affiliation{Department of Applied Physics, Yale University, New Haven, Connecticut 06520, USA}
\affiliation{Yale Quantum Institute, Yale University, New Haven, Connecticut 06520, USA}
\author{Joseph Tindall}
\affiliation{Center for Computational Quantum Physics, The Flatiron Institute, 162 5th Avenue, New York, New York 10010, USA}

\begin{abstract}
Mutually unbiased bases (MUBs) describe quantum measurements for which certainty in one basis gives uniform outcomes in the others. Their maximum number remains unknown in dimensions that are not powers of a prime.
We introduce an ansatz that, in many dimensions, allows us to construct more MUBs than have previously been found. 
Each basis in the ansatz consists of a diagonal phase matrix applied to a fixed unitary, which is a tensor product of Fourier matrices and, optionally, any real Hadamard matrix.
This construction yields five MUBs in dimension 12, six in dimensions 48, 96, and 192, seven in dimensions 36 and 108, and ten in dimension 648.
These bases can then be extended to exceed the tensor product bound, asymptotically, in every eighteenth dimension.
Moreover, we show that Paley's real Hadamard matrix can be used to construct $q+1$ bases in dimension $d = q(q+1)$ for every prime power $q\equiv3\pmod4$. This count grows as $\sqrt d$ and cannot be exceeded by tensor products of smaller sets.
\end{abstract}

\maketitle

\section{Introduction}\label{sec:introduction}
Mutually unbiased bases (MUBs) describe quantum measurements where certainty in one basis gives no information about the other. 
For $d\ge2$, two orthonormal bases, $\mathcal A=\{\ket{a_j}\}_{j=1}^{d}$ and $\mathcal B=\{\ket{b_k}\}_{k=1}^{d}$ in $\C^d$ are mutually unbiased if
\begin{equation}
 |\langle a_j|b_k\rangle|^2=\frac1d
 \quad\text{for all }j,k.
 \label{eq:mub}
\end{equation}
Define the corresponding unitary matrices $A$ and $B$ with columns $|a_j\rangle$ and $|b_k\rangle$, respectively, so that $(A^\dagger B)_{jk}=\langle a_j|b_k\rangle$. The bases are mutually unbiased exactly when $H=\sqrt d\,A^\dagger B$ only has entries of magnitude one.
Since unitarity of $A$  and $B$ also gives $H^\dagger H=dI_d$, where $I_d$ is the $d \times d$ identity matrix, this is precisely the condition that $H$ is a complex Hadamard matrix~\cite{TadejZyczkowski2006,Review2026}. A set of bases is mutually unbiased if every pair of bases is mutually unbiased.

MUBs arose from Schwinger's work on unitary operator bases and Ivanovi\'c's work on quantum state determination~\cite{Schwinger1960,Ivanovic1981}. Wootters and Fields showed how they allow efficient state reconstruction, since each additional basis supplies $d-1$ independent probabilities~\cite{WoottersFields1989}.
This complementarity also underlies uncertainty relations~\cite{Wu2009}, quantum cryptography~\cite{Bruss1998}, and entanglement tests~\cite{Spengler2012,Morelli2023}. Experiments frequently utilize MUBs for quantum cryptography~\cite{Mirhosseini2015}, tomography of entangled photons~\cite{Giovannini2013}, and entanglement certification~\cite{HerreraValencia2020}.

Let $N(d)$ be the size of the largest possible set of MUBs in dimension $d$. The universal upper bound is $N(d)\le d+1$, and equality holds in every \textit{prime power} dimension, i.e. every dimension which satisfies $d= p^a$, where $p$ is prime and $a \in \mathbb N$. A set of $d+1$ MUBs in dimension $d$ is called complete~\cite{WoottersFields1989,Bandyopadhyay2002}. 
Write $d=\prod_{i=1}^s p_i^{a_i}$, where the
$p_i$ are distinct primes, $a_i\ge1$, and $s\ge1$.
Tensoring complete MUBs in each of the prime-power factors gives the tensor product bound~\cite{Klappenecker2004},
\begin{equation}
 N(d)\ge T(d)=1+\min_{i}p_i^{a_i},
 \label{eq:tensor}
\end{equation}
where the factor with the fewest MUBs limits this lower bound. 

This bound can be exceeded. Wocjan and Beth used mutually orthogonal Latin squares to obtain larger sets~\cite{WocjanBeth2005} in certain square number dimensions.
Specific designs give six bases in $d=196$ and seven in $d=324$, exceeding $T(d)=5$~\cite{Todorov2012,Abel2015}. Their constructive method also gives an unbounded gain over $T(d)$ along suitable square dimensions as $d \rightarrow \infty $~\cite{WocjanBeth2005}.

The exact value of $N(d)$ is unknown for all non-prime power dimensions~\cite{Review2026}. The existence of four MUBs in dimension six, for example, is an outstanding open problem in quantum information science ~\cite{Bengtsson2007,ButterleyHall2007,BrierleyWeigert2008,Jaming2009,Horodecki2022,Review2026,wuttig2026completeclassificationcomplexhadamard}. The next highest unresolved dimensions are ten and twelve.

In this work, we construct MUBs beyond the tensor product bound with a simple ansatz where each basis is just a fixed unitary matrix multiplied by a diagonal phase matrix. The unitary is a tensor product of Fourier matrices and, optionally, a real Hadamard matrix. Constructing MUBs then reduces to finding the appropriate phases for a given factorization of the unitary. 
This ansatz allows us to construct five MUBs in dimension 12, six in dimensions 48, 96, and 192, seven in dimensions 36 and 108, and ten in dimension 648. These sets and their tensor extensions exceed $T(d)$ in at least one dimension in eighteen asymptotically. 
Using any real Hadamard inside our ansatz guarantees a set of MUBs in a higher dimension. A special case is Paley's construction, which can be used to  build $q+1$ MUBs in dimension $q(q+1)$ for every prime power $q\equiv3\pmod4$. This count grows as $\sqrt d$ and cannot be exceeded by ordinary tensor products of smaller MUB sets.

\begin{figure*}[htb!]
\centering
\includegraphics[width=\textwidth]{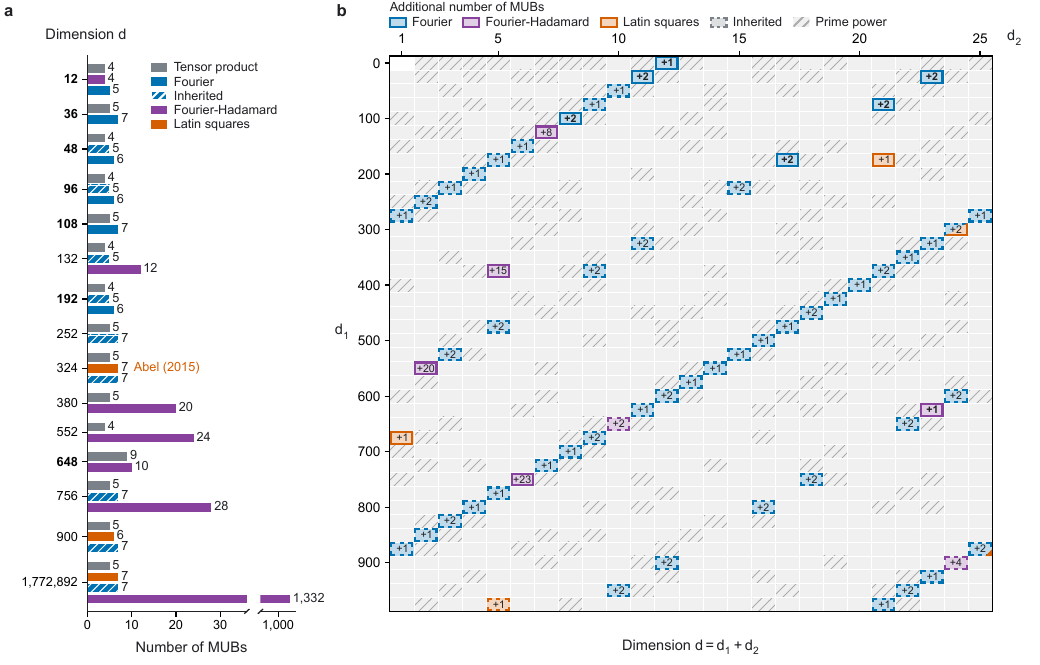}
\caption{\textbf{MUBs beyond the tensor product bound.}
(\textbf{a}) Counts and (\textbf{b}) largest absolute gains over the tensor product bound $T(d)$ from solutions of Eq.~\eqref{eq:condition} using the phased Fourier (blue) or phased Fourier-Hadamard (purple) constructions up to $d=1000$. Each cell corresponds to a dimension $d=d_1+d_2$. All counts immediately give a lower bound on $N(d)$.
Orange shows previously identified Latin-square constructions~\cite{WocjanBeth2005,Todorov2012,Abel2015,Miller2024}. Inherited (dashed outlines) uses tensor products of MUB sets in smaller dimension with the color reflecting the type of sets used.
Previous Latin square constructions match our phased Fourier inherited count at $d=324$ (split box), whilst at $d=900$, our inherited count is larger (orange corner). Our constructions, asymptotically, exceed $T(d)$ in every eighteenth dimension. In prime power dimensions (gray hatched), complete sets of MUBs exist.
}
\label{fig:construction}
\end{figure*}

Figure~\ref{fig:construction} compares the number of MUBs obtained from our ansatz with previously known results. 
Our constructions exceed the tensor product bound in many different dimensions, which prior to this work gave the largest known sets of MUBs in non-prime-power dimensions below 100~\cite{info17080796}.

\section{The phased unitary ansatz}\label{sec:ansatz}
Let $U_d$ be a $d\times d$ unitary whose entries all have magnitude $1/\sqrt d$. We look for bases of the form
\begin{equation}
A_a=\diag(\mathbf w_a)\,U_d,
\label{eq:ansatz}
\end{equation}
where each vector $\mathbf w_a\in\C^d$ has entries of magnitude one, and $\mathbf w_1=(1,\ldots,1)$ always gives $A_1^{(d)} =U_d$.
Every $A_a$ is unitary and unbiased to $I_d$. The set $\{I_d,A_1,\ldots,A_{m-1}\}$ is a set of $m$ MUBs exactly when
\begin{equation}
\left|\left[
U_d^\dagger
\diag\!\left(
\overline{\mathbf w_a}\odot\mathbf w_b
\right)U_d
\right]_{jk}\right|^2=\frac1d
\label{eq:condition}
\end{equation}
for all $a\ne b$ and $j,k$, with $a=1,\ldots,m-1$. Here $\odot$ denotes entrywise multiplication, i.e. the Hadamard product.

We consider two choices of unitaries $U_d$. First, \textit{phased Fourier} constructions for a factorization $d=d_1 \cdot d_2 \cdot ... \cdot d_k$
\begin{equation}
    U_d=F_{d_1}\otimes\cdots\otimes F_{d_k},
\end{equation} 
where $F_d$ is the normalized Fourier matrix with entries
\begin{equation}
    (F_d)_{jk}=e^{2\pi \mathrm{i} (j-1)(k-1)/d}/\sqrt d \qquad 1\le j,k\le d.
\end{equation}
Second, \textit{phased Fourier-Hadamard} constructions 
\begin{equation}
U_d=F_{d_1}\otimes\cdots\otimes F_{d_k}\otimes\frac{H}{\sqrt h},
\label{eq:unitaries}
\end{equation}
where $H$ is a real Hadamard matrix of order $h$, and \hbox{$d=h \cdot d_1 \cdot d_2 \cdot ... \cdot d_k$}.

\subsection{Phased Fourier constructions}
\subsubsection{Five MUBs in dimension 12}
Use $U_{12} = F_6 \otimes F_2$ and define $\zeta=e^{{\rm i}\pi/3}$ with
\begin{align}
\mathbf w_2^{(12)}&=(1,\zeta^2,\zeta^5,\zeta^2,\zeta^2,\zeta^4,\zeta^5,1,1,1,\zeta^5,\zeta^4),\\
\mathbf w_3^{(12)}&=(1,\zeta^5,\zeta^5,\zeta^2,\zeta^4,\zeta^3,\zeta^5,1,1,\zeta^3,\zeta,1),\\
\mathbf w_4^{(12)}&=(1,\zeta^3,\zeta^5,\zeta^5,\zeta^2,\zeta^3,\zeta^3,\zeta^3,\zeta^2,\zeta^5,\zeta^5,\zeta^3).
\end{align}
Then, direct substitution shows that $\sqrt{12}\,(A_a^{(12)})^\dagger A_b^{(12)}$ is a complex Hadamard matrix for $1\le a<b\le4$. Thus, $\{I_{12},A_1^{(12)},A_2^{(12)}, A_3^{(12)}, A_4^{(12)} \}$ gives five MUBs in dimension 12, which exceeds the standard tensor product bound of $T(12) = 4$. 
This construction actually belongs to a continuous family containing, as we show in the Supplemental Material.

\subsubsection{Seven MUBs in dimension 36}
Set $U_{36}=F_3^{\otimes2}\otimes F_2^{\otimes2}$ and $\omega=\zeta^2$. The entries $r = 1,...,36$ of $\mathbf{w}^{(36)}_a$ can be indexed uniquely via
$
    r=1+12x+4z+2y_1+y_2,
$
where $x,z\in\{0,1,2\}$ and $y_1,y_2\in\{0,1\}$. 
At each row, define
\begin{align}
    u=&\omega^{[(z+1)y_1+(z-1)y_2]^2-x(x+z)}\,, \\
    v=&(-1)^{y_1y_2}\omega^{z(z-x)} \,.
\end{align}
Then, the phase vectors are
\begin{equation}
\begin{aligned}
(\mathbf w_2^{(36)})_r&=u, &
(\mathbf w_3^{(36)})_r&=v^{-1}, &
(\mathbf w_4^{(36)})_r&=uv^{-1},\\
(\mathbf w_5^{(36)})_r&=u^2, &
(\mathbf w_6^{(36)})_r&=u^2v.
\end{aligned}
\end{equation}
The set $\{I_{36},A_1^{(36)}, A_2^{(36)}, ..., A_6^{(36)}\}$ then gives seven mutually unbiased bases in $d=36$, exceeding $T(36)=5$. 

\subsubsection{MUBs in dimensions 48, 96, 108 and 192}
We can also use phased Fourier constructions to obtain six MUBs exceeding $T(d) = 4$ in dimensions 48, 96, and 192, which we write as $d=3\cdot2^m$ for $m=4,5,6$, respectively, and take $ U_d=F_3\otimes F_2^{\otimes m}$. In $d=108$, we use $ U_{108}=F_3^{\otimes3}\otimes F_2^{\otimes2}$ to obtain seven MUBs instead of five. The explicit phase vectors are given in the Supplemental Material.

\subsection{Phased Fourier-Hadamard Constructions}
Now we show that any real Hadamard matrix $H$ of order $h \ge 4$ can always be used with $U_d=\tilde{\mathcal F}_{h-1}\otimes H/\sqrt h$ in dimension $d=h(h-1)$ to construct MUBs. Here,
\begin{equation}
      \tilde{\mathcal{F}}_{h-1} := \bigotimes_i F_{p_i}^{\otimes a_i},
\label{eq:FH_unitary}
\end{equation}
and we have used the prime decomposition $h - 1 = \prod_i p_i^{a_i}$. This gives $1 + \lambda(h-1)$ many MUBs, 
where we define $\lambda(n)=\min_i p_i^{a_i}$ for the prime factorization $n=\prod_i p_i^{a_i}$ of any integer $n\ge2$.

Here we'll write down the construction for $s=h-1$ is prime. Then, $\lambda(h-1) = h-1$.
Label the rows of $U_d$ by $(x,y)$, where $x\in\{0,\ldots,s-1\}$ labels a Fourier row and $y\in\{0,\ldots,s\}$ a Hadamard row. The vectors
\begin{equation}
\mathbf w^{(d)}_a(x,y)=
\begin{cases}
e^{2\pi {\rm i}(a-1)(xy-(h/4)y^2)/s},&y<s,\\[2pt]
e^{2\pi {\rm i}(a-1)x^2/s},&y=s,
\end{cases}
\label{eq:hadamard-masks}
\end{equation}
$a=1,\ldots,s$ then solve Eq.~\eqref{eq:condition}.

For composite $s=h-1$, we use one finite field for each full prime-power factor so the exponential evaluated in the phase vector becomes a product over .
each field coordinate for each factor. This gives $\lambda(h-1)$ phase vectors with $\mathbf{w}_1^{(d)}(x,y)$ a vector of ones.
The explicit vectors are given in the Supplemental Material.
We note that related linear and quadratic phases appear in Ref.~\cite{LiYingZhang2026}.

MUBs constructed this way exceed the tensor product bound  $T(h(h-1))=1+\min\{\lambda(h),\lambda(h-1)\}$ exactly when $\lambda(h-1)>\lambda(h)$.
For $h=1$ or $h=2$, we first enlarge $H$ to the real Hadamard matrix $H\otimes[2(F_2\otimes F_2)]$ of order $4h$, then apply the construction.
If the Hadamard conjecture~\cite{Paley1933} is true, this would allow our ansatz to construct MUBs in $d= h(h-1)$ for every order $h$ divisible by four.

An important result can be obtained with our ansatz using Paley's Hadamard matrix of order $q+1$ ~\cite{Paley1933}, which can be constructed for any prime power $q$ where $q\equiv3\pmod4$. For such a $q$ and $d=q(q+1)$, this gives $q+1$ many MUBs. 
This count grows as $\sqrt d$. For example, $q=27$ gives 28 many MUBs in dimension 756, compared with $T(756)=5$, see Fig.~\ref{fig:construction}. Kumar and Maitra constructed $q+1$ approximately unbiased real bases in the same dimensions~\cite{KumarMaitra2022}. The bases we construct here with out ansatz are exactly unbiased.

\subsubsection{Further examples}
In $d=648$, phase vectors built from a published difference matrix~\cite{AbelColbournWojtas2004} give ten MUBs instead of $T(d) = 9$. Write $d=h^2/2$ with $h=4\cdot3^2$, where a real Hadamard matrix $H$ of order $h$ is supplied by Paley's construction~\cite{Paley1933}. Then, $U_{648}=\frac{H}{6}\otimes F_2\otimes\overline{F_3}^{\otimes 2}$ can be used with the phase vectors to exceed the tensor product bound.
In the Supplemental Material, we supply the corresponding phase vectors.

Another interesting example comes from using this ansatz with the recently announced Hadamard matrix of order 668 by Alp\"oge and collaborators~\cite{Alpoge2026Hadamard}. This gives $N(d)\ge24>T(d)=5$ for $d=445\,556=668\cdot667$.

\subsection{Tensor extensions}
Tensoring two sets of MUBs gives $N(de)\ge\min\{N(d),N(e)\}$. Tensoring our constructions here with complete prime-power sets, for example, allows one to construct more MUBs
\begin{equation}
\begin{aligned}
N(d)&\ge6=T(d)+2,\ d=3\cdot2^n,\ n\ge4,\\
N(d)&\ge7=T(d)+2,\ d=4\cdot3^n,\ n\ge2,\\
N(d)&\ge10=T(d)+1,\ d=8\cdot3^n,\ n=4\ \text{or}\ n\ge6.
\end{aligned}
\label{eq:inherited}
\end{equation}
These extensions exceed $T(d)$ throughout the families $d=24n+12$, $d=144n+48$ and $d=144n+96$, where $n \in \mathbb{N}_0$ and hence in at least one dimension in eighteen asymptotically.

Any factorization of a dimension $d = q(q+1)$ into at least two factors greater than one has a factor of dimension at most $q$, so the ordinary tensor product construction cannot exceed $q+1$ bases~\cite{McNulty2016}. This means that our Fourier-Hadamard construction using the Paley matrix with $q\equiv3\pmod4$ cannot be exceeded by tensor products of smaller MUB sets, even if complete sets existed.
Moreover, tensoring any fixed real Hadamard matrix with suitable Paley matrices gives orders $h$ for which $\lambda(h-1)$ grows without bound while $\lambda(h)$ stays bounded, as shown in the Supplemental Material.
Our phased Fourier Hadamard construction then constructs sets of MUBs whose relative size over $T(h(h-1))$ grows without bound.

\section{Experimental implications}
The discovery of new MUBs is useful for a number of experimental tasks. These include reconstructing quantum states~\cite{WoottersFields1989}, detecting entanglement and certifying dimension~\cite{Spengler2012,Morelli2023}, and verifying maximally entangled states as efficiently as possible~\cite{ZhuHayashi2019}.
Certifying high-dimensional entanglement using only a few measurement bases has already been demonstrated with photons encoded in space~\cite{Bavaresco2018,HerreraValencia2020}. In dimension twelve, programmable silicon photonic chips can perform arbitrary local measurements~\cite{Wang2018}. This means the set of five MUBs in dimension $d = 12$ found here, can be tested on existing hardware and used to certify entanglement more robustly.

\section*{Acknowledgements}
M.C.W. acknowledges the hospitality of the Center for Computational Quantum Physics at the Flatiron Institute. The Flatiron Institute is a division of the Simons Foundation.

OpenAI's ChatGPT 6 Astra and Claude Fable 5.1 were prompted to construct mutually unbiased bases that had previously not been found. 
The authors simplified and summarized the results, pushed the LLMs to identify a universal construction from any real Hadamard, independently checked all LLM-assisted material, and verified all claims. The authors take full responsibility for the content and presentation of this work.

\section*{Author contribution}
M.C.W. and J.T. contributed to the conception of the study, the development and verification of the mathematical results, and the preparation of the manuscript. Both authors reviewed and approved the final manuscript.

\bibliography{ref}

\clearpage
\onecolumngrid
\setcounter{secnumdepth}{2}

\setcounter{section}{0}
\setcounter{subsection}{0}
\renewcommand{\thesection}{\Roman{section}}

\setcounter{equation}{0}
\numberwithin{equation}{section}
\renewcommand{\theequation}{S.\arabic{section}.\arabic{equation}}

\setcounter{theorem}{0}
\renewcommand{\thetheorem}{S.\arabic{theorem}}

\makeatletter
\renewcommand{\theHsection}{supp.\arabic{section}}
\renewcommand{\theHequation}{supp.\arabic{section}.\arabic{equation}}
\providecommand{\theHtheorem}{}
\renewcommand{\theHtheorem}{supp.\arabic{theorem}}
\makeatother

\setcounter{table}{0}
\setcounter{figure}{0}
\renewcommand{\thetable}{S\arabic{table}}
\renewcommand{\thefigure}{S\arabic{figure}}
\makeatletter
\renewcommand{\theHtable}{supp.\arabic{table}}
\renewcommand{\theHfigure}{supp.\arabic{figure}}
\renewcommand{\p@subsection}{\thesection.}
\makeatother

\section{Further explicit MUB constructions}\label{supp:explicit}
\subsection{Continuous family of five MUBs in dimension 12}
Use the definitions of $\{A_1^{(12)}, ..., A_4^{(12)}\}$ introduced in the main text for $d = 12$ with the corresponding phase vectors which give 5 MUBS in $d = 12$.
Now, for $\theta\in\mathbb R$, define
\begin{equation}
\bigl(A_a^{(12)}(\theta)\bigr)_{jk}
= e^{{\rm i}\theta\chi_a(j,k)}\,\bigl(A_a^{(12)}\bigr)_{jk},
\end{equation}
with
\begin{equation}
\chi_a(j,k)=
\begin{cases}
1, & j\in\{2,7\} \text{ and } k\in C_a,\\
0, & \text{otherwise},
\end{cases}
\end{equation}
where $C_a=\{2,3,6,7,10,11\}$ for $a\in\{1,3,4\}$ and $C_2=\{1,4,5,8,9,12\}$. Keep the coordinate basis $I_{12}$ fixed.

Rows 2 and 7 of each $A_a^{(12)}$ are equal outside $C_a$ and opposite inside it. Hence
\begin{equation}
A_a^{(12)}(\theta)=
\bigl[I_{12}+(e^{i\theta}-1)vv^\dagger\bigr]A_a^{(12)},
\end{equation}
with $v=\frac{e_2-e_7}{\sqrt2}$, where $e_j$ are coordinate vectors (i.e. they are one in position $j$ and zero elsewhere). The bracket is a unitary matrix, so overlaps between the four bases are unchanged. Their entry magnitudes are also unchanged, preserving unbiasedness to $I_{12}$ and thus $\{I_12, A_1^{(12)}(\theta), ..., A_4^{(12)}(\theta)\}$ is a continuous family of MUBs for $\theta \in \mathbb{R}$ .

\subsection{Six MUBs in dimensions 48, 96, and 192}
Write $d=3\cdot2^m$, with $m=4,5,6$ for dimensions 48, 96, and 192, respectively, and take $U_d=F_3\otimes F_2^{\otimes m}$. A row has index
\begin{equation}
r=1+2^m z+\sum_{j=1}^{m}2^{m-j}y_j,
\label{supp:eq:binary-index}
\end{equation}
with $z\in\{0,1,2\}$ and $ y_j\in\{0,1\}$.
The coordinate $z$ selects one of three row blocks of $U_d$, and the bits $y_j$ give the binary position within that block, with $y_m$ varying fastest.
At $d=48$, the four nonconstant vectors are
\begin{equation}
\begin{aligned}
(\mathbf w_2^{(48)})_r
&=(-1)^{y_3y_4}
  (\omega^{(y_1-y_2)^2},\omega^{y_1},\omega^{1+y_2})_{z+1},\\
(\mathbf w_3^{(48)})_r
&=(-1)^{y_1(y_2+y_4)+y_2y_3}
  (\omega^2,\omega^{(y_2-y_4)^2},\omega^{y_2-y_4})_{z+1},\\
(\mathbf w_4^{(48)})_r
&=(-1)^{y_1(y_2+y_3+y_4)}
  (\omega^{(y_2-y_3)^2},\omega^{-y_3},\omega^{y_2-1})_{z+1},\\
(\mathbf w_5^{(48)})_r
&=(-1)^{(y_1+y_2)y_3+y_2y_4}
  (\omega^{(y_1-y_2)^2-1},1,\omega^{y_2-y_1})_{z+1}.
\end{aligned}
\label{supp:eq:48}
\end{equation}
The subscript $z+1$ selects one entry of the triple and $\omega = e^{{\rm i}2\pi/3}$.

For $d=96$, use Eq.~\eqref{supp:eq:binary-index} with $m=5$ and set
\begin{equation}
(\mathbf w_a^{(96)})_r=(\mathbf w_a^{(48)})_{\lceil r/2\rceil}c_a,
\end{equation}
for $a=2,3,4,5$, where
\begin{equation}
\begin{aligned}
c_2&=i^{y_3}(-1)^{y_1y_4+y_5(y_1+y_3+y_4)},\\
c_3&=i^{y_2+y_5}(-1)^{y_4(y_1+y_2+y_3)},\\
c_4&=i^{y_1+y_2+y_5}(-1)^{y_1(y_3+y_4)+y_5(y_2+y_4)},\\
c_5&=i^{y_1+y_4}(-1)^{y_1y_3+y_5(y_1+y_2)}.
\end{aligned}
\label{supp:eq:96}
\end{equation}
The index $\lceil r/2\rceil$ removes the last bit.

For $d=192$, add a sixth bit $y_6$ and set
\begin{equation}
(\mathbf w_a^{(192)})_r
=(\mathbf w_a^{(96)})_{\lceil r/2\rceil}(-1)^{y_6L_a},
\label{supp:eq:192}
\end{equation}
with
\begin{equation}
    (L_1,L_2,L_3,L_4,L_5)
=(0,y_3,y_2+y_5,y_1+y_2+y_5,y_1+y_4).
\end{equation}

\subsection{Seven MUBs in dimension 108}
Take $U_{108}=F_3^{\otimes3}\otimes F_2^{\otimes2}$ and write the row index as
\begin{equation}
r=1+36x+12z+4t+2y_1+y_2,
\end{equation}
with $x,z,t\in\{0,1,2\}$ and $y_1,y_2\in\{0,1\}$.
Take $u$ and $v$ from the dimension-36 construction in the main text. Then, the six phase-vector entries are
\begin{equation}
\bigl((\mathbf w_1^{(108)})_r,\ldots,(\mathbf w_6^{(108)})_r\bigr)
=\bigl(1,u\omega^{t^2},v^{-1}\omega^{(t-z)^2},
uv^{-1}\omega^{-t^2},u^2\omega^{-(t-z)^2},
u^2v\omega^{-(t+x)^2}\bigr).
\label{supp:eq:108}
\end{equation}

\subsection{Ten MUBs in dimension 648}\label{supp:quotient}
In dimension 648, we use
\begin{equation}
U_{648}=\frac H6\otimes F_2\otimes\overline{F_3}^{\otimes2},
\end{equation}
with $H$ any real Hadamard matrix of order $36$ from Paley's construction~\cite{Paley1933}.
Index the entries of the phase vectors by
\begin{equation}
r=1+18c+9z+3x_1+x_2,
\label{supp:eq:quotient-index}
\end{equation}
with $ c\in\{0,\ldots,35\},\ z\in\{0,1\},\ x_1,x_2\in\{0,1,2\}$,
and set
\begin{equation}
(\mathbf w_a^{(648)})_r={\rm i}^{e_1}\,(-1)^{e_2z}\,\omega^{t_1x_1+t_2x_2},
\label{supp:eq:quotient-masks}
\end{equation}
where $(e_1,e_2,t_1,t_2)$ are the four-digit entries in Table~\ref{supp:tab:dm36}, which is specified by a row $c$ and column $a$. The vector $\mathbf w_1^{(648)}$ is all ones. The table is the difference matrix of Abel, Colbourn, and Wojtas~\cite{AbelColbournWojtas2004}, written out in full. This construction gives ten MUBs.

\begin{table}[htb!]
\centering
\footnotesize\setlength{\tabcolsep}{3pt}
\begin{tabular}{r|cccccccc|r|cccccccc}
\toprule
$c\backslash a$&2&3&4&5&6&7&8&9&$c\backslash a$&2&3&4&5&6&7&8&9\\
\midrule
0&00\,00&00\,00&00\,00&00\,00&00\,00&00\,00&00\,00&00\,00&18&10\,11&11\,01&10\,01&01\,02&11\,00&01\,20&00\,01&00\,02\\
1&01\,00&11\,12&00\,10&01\,20&01\,10&10\,10&11\,00&10\,00&19&11\,11&01\,01&01\,21&00\,22&10\,10&01\,22&11\,22&10\,11\\
2&10\,00&00\,21&10\,10&00\,10&01\,20&11\,01&01\,10&11\,10&20&00\,12&11\,00&11\,21&01\,12&10\,12&11\,11&10\,20&01\,10\\
3&11\,00&00\,12&01\,00&11\,10&11\,20&10\,12&10\,21&01\,12&21&01\,12&10\,12&00\,21&11\,02&10\,22&01\,02&01\,00&11\,20\\
4&00\,01&01\,20&11\,00&10\,20&11\,02&10\,22&01\,02&11\,21&22&10\,12&10\,21&11\,11&11\,22&00\,02&00\,01&11\,20&10\,10\\
5&01\,01&01\,02&10\,20&10\,10&00\,12&00\,21&10\,22&01\,11&23&11\,12&10\,00&01\,11&10\,02&01\,22&11\,10&00\,11&00\,22\\
6&10\,01&11\,11&10\,00&01\,00&11\,22&01\,01&00\,22&00\,11&24&00\,20&00\,10&00\,02&00\,01&00\,22&00\,11&00\,21&00\,12\\
7&11\,01&01\,11&01\,20&00\,20&10\,02&01\,00&11\,10&10\,20&25&01\,20&11\,22&00\,12&01\,21&01\,02&10\,21&11\,21&10\,12\\
8&00\,02&11\,10&11\,20&01\,10&10\,01&11\,22&10\,11&01\,22&26&10\,20&00\,01&10\,12&00\,11&01\,12&11\,12&01\,01&11\,22\\
9&01\,02&10\,22&00\,20&11\,00&10\,11&01\,10&01\,21&11\,02&27&11\,20&00\,22&01\,02&11\,11&11\,12&10\,20&10\,12&01\,21\\
10&10\,02&10\,01&11\,10&11\,20&00\,21&00\,12&11\,11&10\,22&28&00\,21&01\,00&11\,02&10\,21&11\,21&10\,00&01\,20&11\,00\\
11&11\,02&10\,10&01\,10&10\,00&01\,11&11\,21&00\,02&00\,01&29&01\,21&01\,12&10\,22&10\,11&00\,01&00\,02&10\,10&01\,20\\
12&00\,10&00\,20&00\,01&00\,02&00\,11&00\,22&00\,12&00\,21&30&10\,21&11\,21&10\,02&01\,01&11\,11&01\,12&00\,10&00\,20\\
13&01\,10&11\,02&00\,11&01\,22&01\,21&10\,02&11\,12&10\,21&31&11\,21&01\,21&01\,22&00\,21&10\,21&01\,11&11\,01&10\,02\\
14&10\,10&00\,11&10\,11&00\,12&01\,01&11\,20&01\,22&11\,01&32&00\,22&11\,20&11\,22&01\,11&10\,20&11\,00&10\,02&01\,01\\
15&11\,10&00\,02&01\,01&11\,12&11\,01&10\,01&10\,00&01\,00&33&01\,22&10\,02&00\,22&11\,01&10\,00&01\,21&01\,12&11\,11\\
16&00\,11&01\,10&11\,01&10\,22&11\,10&10\,11&01\,11&11\,12&34&10\,22&10\,11&11\,12&11\,21&00\,10&00\,20&11\,02&10\,01\\
17&01\,11&01\,22&10\,21&10\,12&00\,20&00\,10&10\,01&01\,02&35&11\,22&10\,20&01\,12&10\,01&01\,00&11\,02&00\,20&00\,10\\
\bottomrule
\end{tabular}
\caption{Digits $(e_1,e_2,t_1,t_2)$ of $\mathbf w_a^{(648)}$ in Hadamard block $c$. Column $a=1$ is all zeros.}
\label{supp:tab:dm36}
\end{table}

\section{General Hadamard-to-MUB construction}\label{supp:transfer}
We provide the Fourier--Hadamard construction of the main text for every order $h$ divisible by four, including composite $h-1$. The main text gives the construction in the special case where $h-1$ is prime.
Recall that $\lambda(n)=\min_ip_i^{a_i}$ for $n=\prod_ip_i^{a_i}$.

Let $H$ be a real Hadamard matrix of order $h\ge4$. Write $s=h-1=q_1 \cdot q_2 \cdot ... \cdot q_n$, where $q_i=p_i^{k_i}$ are powers $n$ many of distinct primes, and let $\lambda(s)=\min_iq_i$. Take
\begin{equation}
U_d=\tilde{\mathcal F}_s\otimes\frac H{\sqrt h},\qquad \tilde{\mathcal F}_s=\bigotimes_{i=1}^nF_{p_i}^{\otimes k_i},
\end{equation}
in dimension $d=sh$. With $Q_i=q_{i+1}\cdot ... \cdot q_n$ (so $Q_n=1$), index the rows of $U_d$ by
\begin{equation}
r=1+h\sum_{i=1}^nQ_ix_i+Y,\qquad x_i\in\{0,\ldots,q_i-1\},\quad Y\in\{0,\ldots,s\},
\end{equation}
where for $Y<s$ we write $Y=\sum_{i=1}^nQ_iy_i$ with $y_i\in\{0,\ldots,q_i-1\}$. The phase vectors are
\begin{equation}
(\mathbf w_a)_r=
\begin{cases}
\prod_{i=1}^n\exp\!\Bigl[\frac{2\pi{\rm i}}{p_i}(a-1)\bigl(x_iy_i-\frac h4y_i^2\bigr)\Bigr],&Y<s,\\[4pt]
\prod_{i=1}^n\exp\!\Bigl[\frac{2\pi{\rm i}}{p_i}(a-1)x_i^2\Bigr],&Y=s,
\end{cases}
\label{supp:eq:transfer-masks}
\end{equation}
with $a=1,\ldots,\lambda(s)$. Together with $I_d$, the bases $A_a=\diag(\mathbf w_a)U_d$ form $1+\lambda(s)$ many MUBs. For prime $s$, this is the formula of the main text.

Note that if some $q_i=p^k$ is a prime power with $k>1$, the numbers $x_i$, $y_i$ and $(a-1)$ in the $i$-th factor of Eq.~\eqref{supp:eq:transfer-masks} are replaced by $k\times k$ integer matrices. Specifically, write $x_i$ in base $p$ as $x_i=x_{i,0}+p\,x_{i,1}+\cdots+p^{k-1}x_{i,k-1}$, and replace it $x_i \to M_{x_i}$ by
\begin{equation}
M_{x_i}=x_{i,0}I+x_{i,1}C+\cdots+x_{i,k-1}C^{k-1}
\end{equation}
for an appropriately chosen integer matrix $C$, which can always be achieved. Replace $y_i \to M_{y_i}$ and $(a-1) \to M_{a-1}$ in the same way. The expression in the $i$-th exponent is then evaluated with ordinary matrix products, and the square bracket in Eq.~\ref{supp:eq:transfer-masks} extracts its top-left entry. Thus, for a matrix $M$, the expression means $[M] = M_{11}$.

For $q_i=27$, for example, we can take
\begin{equation}
C=\begin{pmatrix}0&0&-1\\1&0&1\\0&1&0\end{pmatrix},
\end{equation}
and for $q_i=9$ we can take $C=\bigl(\begin{smallmatrix}0&-1\\1&0\end{smallmatrix}\bigr)$. For a prime $q_i$, $C=1$ and Eq.~\eqref{supp:eq:transfer-masks} is used verbatim.
\section{Unbounded advantage from any real Hadamard matrix}\label{supp:amplifier}

Start with a real Hadamard matrix $H_0$ of order $h_0$, and fix a prime $r>2$ that does not divide $h_0$. For any $L\ge r$, choose a prime $q\equiv3\pmod4$ such that $q+1$ is divisible by every prime up to $L$, but not by $r^2$. Such primes exist by the Chinese remainder theorem and Dirichlet's theorem. Let $H_q$ be Paley's matrix of order $q+1$~\cite{Paley1933}. Then $H_0\otimes H_q$ is a real Hadamard matrix of order $h=h_0(q+1)$.

Every prime up to $L$ divides $h$, so none divides $h-1$. Applying our phased Fourier-Hadamard construction using $H = H_0\otimes H_q$ therefore gives $1+\lambda(h-1)>L+1$ MUBs. Meanwhile, $r$ divides $h$ exactly once, so the tensor product bound is at most $r+1$. This gives
\begin{equation}
\frac{1+\lambda(h-1)}{T(h(h-1))}>\frac{L+1}{r+1}.
\end{equation}
Since $r$ is fixed and $L$ can be arbitrarily large, the multiplicative advantage is unbounded.

\end{document}